\documentclass[aps,prl,reprint,superscriptaddress]{revtex4-1}

\usepackage{amsmath,amssymb,bm}
\usepackage{graphicx}
\usepackage{hyperref}
\usepackage{float}
\usepackage{braket}
\usepackage{color}

\renewcommand\Re{{\rm Re}}
\renewcommand\Im{{\rm Im}}

\begin{document}

\title{Macroscopic Quantum Violation of Fluctuation-Dissipation Theorem in Equilibrium}
\author{Kentaro Kubo}
\email{kubo@ASone.c.u-tokyo.ac.jp}
\affiliation{Department of Basic Science, University of Tokyo, 3-8-1 Komaba, Meguro, Tokyo 153-8902, Japan}
\author{Kenichi Asano}
\email{asano@phys.sci.osaka-u.ac.jp}
\affiliation{Department of Physics, Osaka University, Toyonaka, Osaka 560-0043, Japan}
\author{Akira Shimizu}
\email{shmz@ASone.c.u-tokyo.ac.jp}
\affiliation{Department of Basic Science, University of Tokyo, 3-8-1 Komaba, Meguro, Tokyo 153-8902, Japan}

\date{\today} 

\begin{abstract}
We examine the Hall conductivity of macroscopic two-dimensional quantum system, and show that the observed quantities can sometimes violate the fluctuation dissipation theorem (FDT), even in the linear response (LR) regime infinitesimally close to equilibrium.
The violation can be an order of magnitude larger than the Hall conductivity itself  at low temperature and in strong magnetic field, which are accessible in experiments.
We further extend the results to general systems and give a necessary condition for such large-scale violation to happen. 
This violation is a genuine quantum phenomenon that appears on a macroscopic scale.
Our results  are not only bound to the development of the fundamental issues of nonequilibrium physics, but the idea is also meaningful 
for practical applications, since the FDT is widely used for the estimation of noises from the LRs.
\end{abstract}

\maketitle

{\em Introduction.---}
According to the fluctuation dissipation theorem (FDT), fluctuation in a thermal equilibrium state should agree with the linear response (LR) function times temperature $T=1/\beta$ (we take $k_{\rm B}=1$) \cite{Einstein, Johnson, Nyquist, Landaustatphys, Takahashi, CallenWelton, Nakano, Kubo1, Kubo2,KTH,FS2016,SF2017}.
While the FDT was originally proposed 
on the dissipative components of LRs, such as the diagonal conductivity \cite{Einstein, Johnson, Nyquist}, it was later suggested  \cite{Onsager1, Onsager2} and proved  \cite{Takahashi} in classical systems that the FDT holds also for the dissipationless components of LRs, such as the Hall conductivity.
Consequently, the FDT is accepted as a universal relation that holds for {\em all} LRs in classical systems.
This contrasts with the responses to strong external fields that drive the systems far away from equilibrium, for which such a relation does not hold in general
 \cite{sn1,sn2,sn3,SU1992,SUS1993,YS2009,violationofness1, violationofness2, violationofness3, violationoffrustratedisinglatticegas, violationofglass, violationofglassformingliquid, violationofsperspinglass, violationofglassreview,SY2010,S2010}. 

The most important aspect of the FDT is that it connects the results of completely different and independent experiments
 \cite{Einstein, Johnson, Nyquist, Landaustatphys, Takahashi,CallenWelton, Nakano, Kubo1, Kubo2,KTH,FS2016,SF2017}:
The LR function is obtained by measuring the response in a nonequilibrium state, whereas the fluctuation is obtained by measuring the time correlation in an equilibrium state.
According to the FDT, one can tell the magnitude of the equilibrium fluctuation by measuring the response function in nonequilibrium, and vise versa.
This nontrivial aspect of the FDT is utilized widely, e.g., to estimate noises (fluctuations) from the responses in electric circuits \cite{electricalengineering1,electricalengineering2,ex3}, optical devices  \cite{ex1,ex2,ex3}, and gravitational-wave detectors  \cite{ex4}.

However, the validity of the FDT is nontrivial in quantum systems 
 \cite{Takahashi,Gardiner,FS2016,SF2017}.
In  ``deriving'' the FDT microscopically 
 \cite{CallenWelton, Nakano, Kubo1, Kubo2}, it was (implicitly) assumed that the disturbance by measurement were negligible. 
Such an ideal measurement is possible only for classical systems, 
because quantum systems follow the uncertainty relations 
\cite{Glauber, MandelWolf, KoshinoShimizu, WisemanMilburn, NielsenChuang}.
The best solution seems to assume a quantum measurement as {\em quasiclassical}, emulating the ideal classical one as closely as possible
 \cite{FS2016,SF2017}.
Then, a question arises: does the FDT hold in quantum systems if  quasiclassical measurements are made ?

This fundamental question was recently been solved in Refs.~\cite{FS2016, SF2017}.
It was shown rigorously that the FDT, as a relation between the results of two different experiments, is partially violated in quantum systems.
Unlike obvious violations far from equilibrium \cite{sn1,sn2,sn3,SU1992,SUS1993,YS2009,violationofness1, violationofness2, violationofness3, violationoffrustratedisinglatticegas, violationofglass, violationofglassformingliquid, violationofsperspinglass, violationofglassreview,SY2010,S2010}, this violation occurs between equilibrium fluctuations and the LRs to infinitesimal external fields.
This violation should also be distinguished 
from that caused at high frequencies $\hbar\omega\gtrsim T$ by 
the {\em non}-quasiclassical detectors which
cannot measure zero-point fluctuations 
\cite{Koch1982,Gardiner,ec1,ec2,ec3,ec4,ec5,ec6,ec7}. 
In fact, the violation does occur 
for certain responses even at $\omega=0$ and even in the quasiclassical measurement by means of the heterodyning  \cite{Koch1982} or quantum nondemolition \cite{PSJmeeting,IQEC,QNDpra,QND1,QND2} detectors.

However, Refs.~\cite{FS2016, SF2017} did not estimate the {\em magnitude} of the violation in actual systems, whereas they derived its general {\em formula}.
Since the violation is a genuine quantum 
phenomenon that vanishes at $\hbar=0$, one might expect that it would be very small in macroscopic systems.
For example, according to Nyquist  \cite{Nyquist}, a macroscopic phenomenology holds using the observed LR functions as given parameters, 
although their values may be determined by quantum effects.
Then the remaining quantum correction, 
for the fluctuation and the LR at frequency $\omega$, 
should be only about the equipartition law, 
which breaks down only when $\hbar\omega\gtrsim T$  \cite{Nyquist}.
This suggests that at low frequency, $\hbar \omega \ll T$, the violation would be irrelevant to macroscopic systems.

In this paper, we explicitly calculate the magnitude of the violation for the Hall conductivity $\sigma_{xy}$ of a macroscopic quantum system.
We show that the violation can be larger by an order of magnitude than $|\sigma_{xy}|$ even at $\omega=0$, in contrast to the above conjecture.
We also derive a condition for such a macroscopic violation of the FDT in general quantum systems.
These results imply that 
the foundation of nonequilibrium physics
needs to be updated even in the LR regime 
infinitesimally close to equilibrium.
Furthermore, our results also have an impact 
on practical applications because
the standard technique of estimating equilibrium noises of currents
is to estimate them from the LR functions using the FDT 
 \cite{electricalengineering1,electricalengineering2,ex1,ex3,ex2,ex4},
but 
wrong predictions are obtained about the `real antisymmetric parts' (see below), such as the Hall conductivity.

{\em FDT violation in Hall effect at $\omega=0$.---}
Consider a macroscopic two-dimensional electron system (2DES) in a static and uniform magnetic flux density, $\bm{B} =(0,0,B)$, which was studied extensively regarding the quantum Hall effect (QHE) \cite{vonKlitzing,wakabayashikawaji,wakabayashikawaji2, SCBA, expofsigmaxx, Ando, TKNN, laughlinQHE, Halperinedge, tsui, laughlinFQHE, HaldaneHierarchicalstructure, HalperinHierarchicalstructure, stormertsui, recent1, recent2}.
We study the FDT between $\sigma_{xy}$ and the time correlation $\Xi_{xy}(t)$ of the current densities $\hat{j}_{x}$ and $\hat{j}_{y}$.

The following were proved
 \cite{FS2016,SF2017} for the case where these quantities are measured quasiclassically, i.e., in such a way that it emulates the classical ideal one as closely as possible: 
As the disturbance on $\sigma_{xy}$ by the measurement is negligibly small, the observed $\sigma_{xy}$ should agree with the one calculated by the Kubo formula, 
although the formula was derived neglecting the disturbance 
 \cite{Kubo1, Kubo2,KTH}.
By contrast, the disturbance on $\Xi_{xy}(t)$ is significantly large, and what one observes is the symmetrized time correlation
 \cite{FS2016,SF2017}, $\Xi_{xy}(t)=\langle\mbox{$\frac{1}{2}$}\{\hat{j}_y(0) \hat{j}_x(t)+\hat{j}_x(t) \hat{j}_y(0)\}\rangle_{\rm eq}$, which differs from the time correlation of the Kubo formula.
[Here, $\langle \bullet \rangle_{\rm eq}$ denotes the equilibrium expectation value.]
This is a universal result that is independent of details of 
the quasiclassical measurement \cite{FS2016,SF2017}.
Consequently, the FDT, relating the observed $\Xi_{xy}$ and $\sigma_{xy}$, is violated.

The violation is quantified as follows; 
For simplicity, we assume that the system is invariant under rotation by $\pi/2$ about the $z$ axis.
Then, the Hall conductivity has only the antisymmetric part: 
$\sigma_{xy}=\sigma^{-}_{xy}:=(\sigma_{xy}-\sigma_{yx})/2$  \cite{KTH}.
From the observed time correlation $\Xi_{xy}(t)$, 
we define 
\begin{align}
S_{xy}(\omega) &:= \int_{0}^{\infty} \Xi_{xy}(t) e^{i\omega t}dt,
\label{eq:S}
\\
\tilde{S}_{xy}(\omega) &:= \int_{-\infty}^{\infty}\Xi_{xy}(t)e^{i\omega t}dt,
\label{eq:tildeS}
\end{align}
which are different only in the lower limit of the integral.
Note that if $\tilde{S}_{xy}$ were employed as `fluctuation', the FDT would be violated, $\sigma_{xy} \neq \beta \tilde{S}_{xy}$, 
even in classical systems  \cite{SF2017}. 
To avoid such superficial violation caused by the improper choice of fluctuation in the frequency domain, 
we instead employ $S_{xy}$ as fluctuation.
Still, in quantum systems, the violation of FDT occurs 
even at $\omega=0$  \cite{FS2016,SF2017}.
To see it clearly, we note that 
$\sigma_{xy}(0)$ satisfies the dispersion relation  \cite{KTH},
\begin{equation}
\sigma_{xy}(0)
=\Re\sigma_{xy}(0)=\int_{-\infty}^{\infty}\frac{\mathcal{P}}{\omega}\Im \sigma_{xy}(\omega)\frac{d\omega}{\pi}, 
\label{eq:KK}
\end{equation}
while the relations  \cite{FS2016,SF2017,KTH} 
\begin{align}
&S_{xy}(0)
={\rm Re} S_{xy}(0)
=
\int_{-\infty}^{+\infty}
{\mathcal{P} \over \omega}
\Im \tilde S_{xy}(\omega)
{d\omega \over 2\pi},
\label{eq:S(0)}
\\
&\beta\Im\tilde{S}_{xy}(\omega)
=
2I_{\beta}(\omega)\Im \sigma_{xy}(\omega)
\label{eq:tildeImS}
\end{align}
%
%
lead to
\begin{equation}
\beta S_{xy}(0)=\int_{-\infty}^{\infty}\frac{\mathcal{P}}{\omega}I_{\beta}(\omega)\Im \sigma_{xy}(\omega)\frac{d\omega}{\pi}.
\label{eq:betaS}
\end{equation}
Here, $\mathcal{P}$ denotes the principal value, and 
\begin{equation}
\!\!
I_{\beta}(\omega)
:=
\frac{\beta\hbar\omega}{2}
\coth \frac{\beta\hbar\omega}{2}
\sim
\begin{cases}
1 & (\beta \hbar |\omega| \ll 1),\\
\beta \hbar |\omega| /2 & (\beta \hbar |\omega| \gg 1).
\end{cases}
\label{eq:I_beta}
\end{equation}
Equations (\ref{eq:KK}) and (\ref{eq:betaS}) show that 
the FDT, 
$\sigma_{xy}(0) = \beta S_{xy}(0)$,
between the observed quantities 
is violated because of the extra factor $I_{\beta}(\omega)$ in 
Eq.~(\ref{eq:betaS}).
Since $I_{\beta}(\omega)=1$ for $\hbar=0$, 
the violation is a genuine quantum phenomenon \cite{FS2016,SF2017}.
We will show that its magnitude can be even {\em larger} than the typical value of $\sigma_{xy}$.

\begin{figure*}[ht]
\centering
\includegraphics[width=\textwidth]{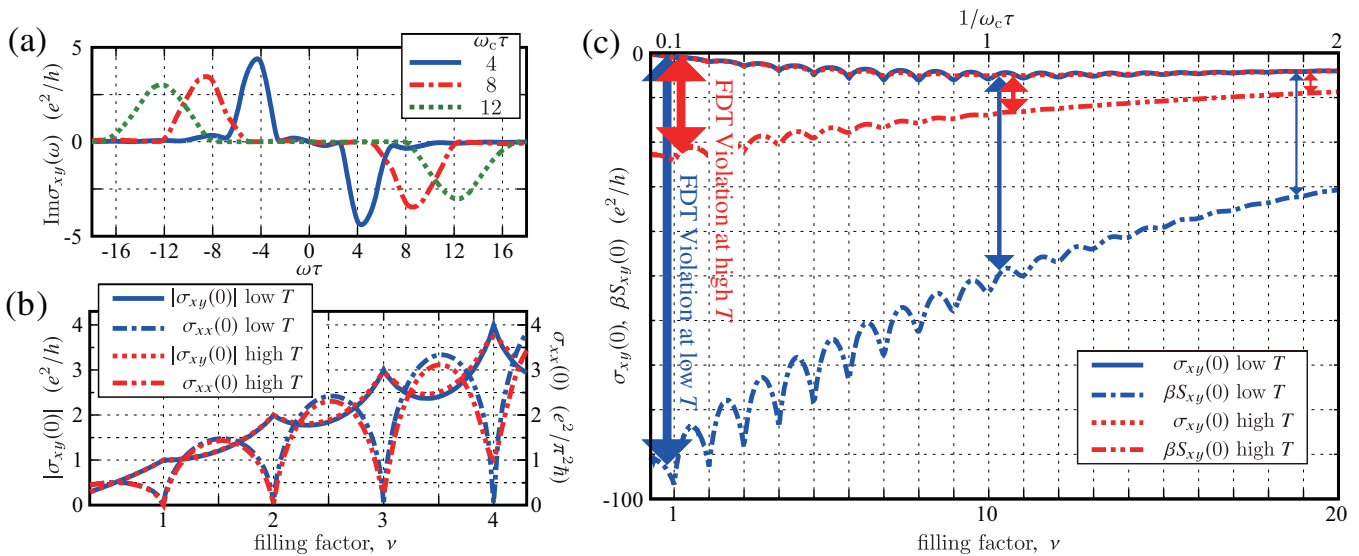}
\caption{ (color online). 
(a) $\Im\sigma_{xy}(\omega)$ as a function of frequency, $\omega$, 
for $\omega_{\rm c}\tau=4$, $8$ and $12$ 
at temperature, $T=\hbar/20\tau$, and electron density, $n=10m/h\tau$. Cyclotron frequency and the scattering time at $B=0$ are denoted as $\omega_{\rm c}$ and $\tau$, respectively.
(b) $|\sigma_{xy}(0)|$ and $\sigma_{xx}(0)$ as a function of filling factor, $\nu$, 
at $T=\hbar/20\tau$ (low $T$) and $T=\hbar/5\tau$ (high $T$) for fixed electron density, $n=10m/h\tau$.
(c) $\sigma_{xy}(0)$ and $\beta S_{xy}(0)$ as a function of $\nu$ for the same values of parameters as in (b).}
\label{fig:all}
\end{figure*}

{\em Conditions for larger violation.---}
Equations (\ref{eq:KK}), (\ref{eq:betaS}) and (\ref{eq:I_beta}) show that the FDT would not be violated if $\Im\sigma_{xy}(\omega)$ had nonzero values only at $\hbar\omega\ll T$.
On the other hand, since the optical absorption (cyclotron resonance) spectra for the left (L) and right (R) circularly polarized light are proportional to the real part of $\sigma^{\rm L/R}_{xy}(\omega):=\sigma_{xx}(\omega)\pm i\sigma_{xy}(\omega)$ \cite{LandauEM}, $\Im\sigma_{xy}(\omega)$ should have positive and negative peaks at around $\omega=-\omega_{\rm c}$ and $+\omega_{\rm c}$, respectively, where
$\omega_{\rm c}:= eB/m$
is the cyclotron frequency.
This is shown in Fig.~\ref{fig:all}(a), which is obtained by the method that we shall explain shortly. 
Therefore, we can expect significant FDT violation when
\begin{equation}
T\lesssim\hbar\omega_{\rm c}.
\label{eq:omegac>T}
\end{equation}
As we will show in {\em General consideration} below, 
the violation is further enhanced, if the spectral peak at around $\omega=\pm\omega_{\rm c}$ is narrow,
\begin{equation}
\hbar\omega_{\rm c}\gtrsim 2\Gamma,
\label{eq:omegacgamma}
\end{equation}
where $\Gamma$ denotes the half width of the Landau level 
which is caused by the impurity scattering \cite{SCBA, Ando}.
We henceforth assume inequalities (\ref{eq:omegac>T}) and (\ref{eq:omegacgamma}) for $T$ and $B$.
The integer QHE can also be expected at sufficiently low temperature, $T\ll\hbar\omega_{\rm c}-2\Gamma$, whereas such extremely low temperature 
(or QHE) is not necessarily required for observing the FDT violation clearly.

{\em Model and method of calculation.---}
To avoid the inconsistency, which could arise due to the separate evaluation at different levels of approximation, we evaluate not only the FDT violation, $\sigma_{xy}(0)-\beta S_{xy}(0)$, but also each of $\sigma_{xy}(0)$ and $\beta S_{xy}(0)$ from a single quantity, $\Im \sigma_{xy}(\omega)$, by means of Eqs.~(\ref{eq:KK}) and (\ref{eq:betaS}).

To calculate $\Im \sigma_{xy}(\omega)$, 
we assume non-interacting electrons
(with a charge $-e$, mass $m$) in two dimension
with the following single-body Hamiltonian:
\begin{align}
\hat{H}
=
\frac{1}{2m}
\left( \hat{\bm{p}}+e\bm{A}(\hat{\bm{r}}) \right)^{2}
+\sum_{i} V_0 \delta(\hat{\bm{r}}-\bm{R}_i).
\label{eq:Hamiltonian}
\end{align}
Here, 
$\bm{A} = (0, Bx)$ is the vector potential, 
$\bm{R}_i$ the location (obeying the uniform distribution) of an impurity, and $V_0$ is the strength of the impurity potential.
The coordinate and momentum operators of the electron are denoted as $\hat{\bm r}$ and $\hat{\bm p}$, respectively.

To avoid artificial divergences caused by subtle treatment of the Landau level degeneracies, we employ the self-consistent Born approximation (SCBA)  \cite{SCBA, Ando}.
The SCBA properly gives the quantized Hall conductivity $\sigma_{xy}(0) = -\nu e^2/h$, when the filling factor 
$\nu:=2\pi l^2n=nh/eB$
is an integer.
Here, $l=\sqrt{\hbar/eB}$ is the magnetic length, and $n$ the electron density.
The formula derived in Ref.~\cite{center}, which is useful to calculate 
$\sigma_{xy}(0)$  \cite{SCBA, Ando}, is inconvenient for our purpose, i.e., to evaluate $\Im \sigma_{xy}(\omega)$ for all $\omega$ including $\omega\sim \pm \omega_{\rm c}$.
We thus perform a straightforward calculation of the Kubo formula using the Landau-level basis and the SCBA.
Then, we can obtain $\Im \sigma_{xy}(\omega)$ by solving the following self-consistent equation for the self-energy $\Sigma(\epsilon)$,
\begin{equation}
\Sigma(\epsilon)=\frac{n_{i}V_{0}^2}{2\pi l^2}\sum_{N}\frac{1}{\epsilon-\hbar\omega_{\rm c}(N+1/2)-\Sigma(\epsilon)}, 
\label{eq:SCBA}
\end{equation}
where $N$ is the Landau index and $n_{i}$ the density of impurities.
We can also estimate the half width of the Landau level, $\Gamma$, from $\Sigma(\epsilon)$.
It has the asymptotic form \cite{SCBA, Ando},
\begin{equation}
\Gamma\sim\sqrt{(2/\pi)(\hbar/\tau)\hbar\omega_{\rm c}},\quad(\omega_{\rm c}\tau\rightarrow\infty), 
\end{equation}
where the scattering time at $B=0$ is denoted as
\begin{equation}
\tau=\left[(2\pi/\hbar)n_{i}V_{0}^2(m/2\pi\hbar^2)\right]^{-1}.
\end{equation}

{\em Diagonal and Hall conductivities.---}
We first confirm that the quantities 
evaluated from $\Im \sigma_{xy}(\omega)$
agree with the ones in the previous works calculated directly without referring to $\Im \sigma_{xy}(\omega)$  \cite{Ando}.

Figure~\ref{fig:all}(b) shows $|\sigma_{xy}(0)|$ ($= -\sigma_{xy}(0)$), obtained from Eq.~(\ref{eq:KK}), as a function of the filling factor $\nu$ at $T=\hbar/20\tau$ and $\hbar/5\tau$. 
The electron density is fixed to $n=10m/h\tau$, and the increase of $\nu$ implies the decrease of $B$.
Inequalities (\ref{eq:omegac>T}) and (\ref{eq:omegacgamma}) hold at $\nu\lesssim 4$.
At low temperature, $T=\hbar/20\tau$, integer QHE is expected.
Indeed, $\sigma_{xy}(0)$ is quantized to $-\nu e^2/h$ at $\nu=1$, $2$, $3$ and $4$, in agreement with the previous theories  \cite{Ando} and experiments  \cite{vonKlitzing, wakabayashikawaji, wakabayashikawaji2}.
At high temperature, $T=\hbar/5\tau$, the quantization blurs at $\nu=3$ and $4$, because thermal excitation to the higher Landau levels takes place.
Since these results are obtained by integrating $\Im \sigma_{xy}(\omega)$, they indicate that $\Im \sigma_{xy}(\omega)$ is reasonably obtained for all $\omega$.

We also calculate the diagonal conductivity $\sigma_{xx}(0)$ as shown in Fig.~\ref{fig:all}(b).
It is nearly quantized as $\sigma_{xx}(0)= (e^2/\pi^2 \hbar)\nu$ for half-integer values of $\nu$, in agreement with the previous studies \cite{wakabayashikawaji,wakabayashikawaji2,SCBA, expofsigmaxx}.
When an external electric field is applied in the $x$-direction, dissipation occurs if $\sigma_{xx} > 0$.
This happens for every $\nu$, except when the QHE takes place at integer-$\nu$ and at low $T$. 

{\em Results for FDT violation.---} In  Fig.~\ref{fig:all}(c), the difference between $\sigma_{xy}(0)$ and $\beta S_{xy}(0)$ 
indicates 
the magnitude of violation of FDT as a function of $\nu$. 
Here, the scale of the vertical axis is about $20$ times larger than that of Fig.~\ref{fig:all}(b).
The magnitude of the violation is enhanced at low temperature, $T=\hbar/20\tau$, which is of an order of magnitude larger than $|\sigma_{xy}(0)|$.
This surprising result should be contrasted with the naive conjecture mentioned in {\em Introduction} that FDT violation would be small in macroscopic systems. 

The violation is enhanced with decreasing $\nu$ and lowering $T$.
This result confirms the expectation in 
{\em Condition for larger violation}
that the FDT is violated significantly when $\hbar \omega_{\rm c} \gtrsim T$.

{\em General consideration.---}
Let us extend our argument to the general macroscopic systems (such as ones with many-body interactions) and to the LRs of the general macroscopic observable of `current' $\dot{\hat{\bm a}}:=d\hat{\bm a}/dt$, associated with the `displacement' $\hat{\bm a}$.
Following Ref.~\cite{SF2017}, we compare the `fluctuation' between two components of $\Delta\dot{\hat{\bm a}}:=\dot{\hat{\bm a}}-\langle\dot{\hat{\bm a}}\rangle_{\rm eq}$, 
\begin{equation}
S_{\mu\nu}(\omega):=\int_0^{\infty}\left\langle\frac 12\{\Delta\dot{\hat{a}}_\mu(0),\Delta\dot{\hat{a}}_\nu(t)\}\right\rangle_{\rm eq}e^{i\omega t}dt.\label{eq:GeneralFluctuation}
\end{equation}
with the admittance tensor, $\chi_{\mu\nu}(\omega)$, describing the LR of $\dot{\hat a}_\mu$ to the external field coupled to $\hat a_\nu$.
When  the symmetric ($+$) and antisymmetric ($-$) parts 
of a tensor $\mathcal{T}$
are defined as 
$\mathcal{T}_{\mu\nu}^{\pm}
:=(\mathcal{T}_{\mu\nu}\pm \mathcal{T}_{\nu\mu})/2$, 
the power loss or the energy dissipation of the external field of frequency, $\omega$, is determined only by $\Re\chi^{+}_{\mu\nu}(\omega)$ and $\Im\chi^{-}_{\mu\nu}(\omega)$, but is independent of $\Im\chi^{+}_{\mu\nu}(\omega)$ and $\Re\chi^{-}_{\mu\nu}(\omega)$ \cite{KTH}. 
\par
In classical systems, 
the FDT, $\chi_{\mu\nu}(\omega)=\beta S_{\mu\nu}(\omega)$, rigorously holds for any choice of $\mu$, $\nu$ and $\omega$ 
\cite{Takahashi,FS2016,SF2017}.
However, in the quantum systems, 
it could be violated regarding the dissipationless component even at $\omega=0$ as $\chi^{-}_{\mu \nu}(0)\ne\beta S^{-}_{\mu \nu}(0)$, 
whereas the dissipative component still 
obeys the FDT as $\chi^{+}_{\mu \nu}(0)=\beta S^{+}_{\mu\nu}(0)$ \cite{FS2016,SF2017}.
 (Note that $\chi^\pm_{\mu\nu}(0)$ and $S^\pm_{\mu\nu}(0)$ are real.)
Actually, Eq.~(60) in Ref.~\cite{SF2017} and Eq.~(4.4.16) in Ref.~\cite{KTH} lead to
\begin{align}
\chi^{-}_{\mu \nu}(0)- \beta S^{-}_{\mu \nu}&(0)
=
\int_{-\infty}^{\infty}\frac{\mathcal{P}}{\omega}\left[1-I_{\beta}(\omega)\right]\Im\chi_{\mu \nu}^{-}(\omega)\frac{d\omega}{\pi}
\nonumber\\
&=
\beta\int_{-\infty}^{\infty}\mathcal P\frac{I_{\beta}(\omega)-1}{\beta\omega}\Re\chi^{\rm L}_{\mu\nu}(\omega)\frac{d\omega}{\pi},
\label{eq:generalViolation} 
\end{align}
where 
\begin{equation}
\chi^{\rm L}_{\mu\nu}(\omega)
:=
[\chi_{\mu\mu}(\omega)+\chi_{\nu\nu}(\omega)]/2+i\chi^{-}_{\mu\nu}(\omega).
\label{eq:chiL}
\end{equation}
Since $\Re \chi^{\rm L}_{\mu\nu}$
should be nonnegative according to the second law
(because it describes the power absorption spectrum of a rotating external field), 
the r.h.s. of Eq.~(\ref{eq:generalViolation}) does not vanish in general.

The magnitude of the FDT violation is roughly
determined by the difference between the spectral intensities of $\Re\chi^{\rm L}_{\mu\nu}(\omega)$ distributed in $\beta\hbar\omega\gtrsim 1$ and in $\beta\hbar\omega\lesssim -1$, because $[I_\beta(\omega)-1]/\beta\omega$ vanishes at $\beta\hbar|\omega|\ll 1$, and can be approximated to the sign function, ${\rm sgn}(\omega)$, at $\beta\hbar|\omega|\gg 1$.
Thus, we can expect significant violation at 
\begin{equation}
T\lesssim\hbar\bar\omega:=\hbar\frac{\left|\int_{-\infty}^{\infty}\omega\Re\chi^{\rm L}_{\mu\nu}(\omega)d\omega/\pi\right|}{\int_{-\infty}^{\infty}\Re\chi^{\rm L}_{\mu\nu}(\omega)d\omega/\pi},\label{eq:GeneralCondition}
\end{equation}
as long as the spectral first moment, $\bar\omega$, is finite.
To evaluate $\bar\omega$, we can use the moment sum rule \cite{KTH}, 
\begin{align}
 \int_{-\infty}^{\infty}\omega^n\Re\chi^{\rm L}_{\mu \nu}(\omega)
\frac{d\omega}\pi
&=
\frac{i(-i)^n}{\hbar}\left\langle\left[\dot{\hat c}_-,\frac{d^n\hat c_+}{dt^n}\right]\right\rangle_{\rm eq},\label{eq:1stMom}
\end{align}
with $n=0,1$ and $\hat c_\pm:=(\hat a_\mu\pm i\hat a_\nu)/\sqrt{2}$.
The violation even diverges 
in proportion to $\beta$ 
at $T\rightarrow 0$ with the asymptotic form,
\begin{equation}
\chi^{-}_{\mu \nu}(0)-\beta S^{-}_{\mu \nu}(0)\sim\beta\int_{-\infty}^{\infty}{\rm sgn}(\omega)\Re\chi^{\rm L}_{\mu\nu}(\omega)\frac{d\omega}{\pi},\label{eq:Asimptotic}
\end{equation}
if the integral in the right hand side is not canceled out.

As for our example of the Hall conductivity, $\hat{\bm a}$ stands for $-e\sum_i\hat{\bm r}_i/\sqrt{S}$, where $S$ and $\hat{\bm r}_i$ are the area of the system and the coordinate operator of each electron, respectively.
Thus, 
inequality (\ref{eq:GeneralCondition}) reproduces (\ref{eq:omegac>T}), since Eq.~(\ref{eq:1stMom}) is explicitly calculated as $ne^2/m$ and $ne^3B/m^2$ for $n=0$ and $1$, respectively. 
Also, Eqs.~(\ref{eq:1stMom}), (\ref{eq:Asimptotic}) and $|\sigma_{xy}(0)|\sim \nu e^2/h=ne/B$ lead to
\begin{equation}
|\sigma^{-}_{xy}(0)-\beta S^{-}_{xy}(0)|/|\sigma_{xy}(0)|\sim\beta\hbar\omega_{\rm c}/2,\quad(T\rightarrow 0)\label{eq:AsimptoticQH}
\end{equation}
in the clean system of $\omega_{\rm c}\tau\gg 1$, 
for which 
${\rm Re}\sigma^{\rm L}_{xy}(\omega)$ almost vanishes at $\omega<0$.
For $n=10m/h\tau$, $\nu=1$ and $T=\hbar/20\tau$, the right hand side of Eq.~(\ref{eq:AsimptoticQH}) equals to $10^2$, which is consistent with our data shown in Fig.~\ref{fig:all}(c).

{\em Notes.---}
Although Fig.~\ref{fig:all} is obtained by SCBA for 
a noninteracting system with short-range impurities, our general argument based on the sum rules is rigorous and independent of the details of the impurities and the electron-electron interaction.
Therefore, the FDT violation in the Hall conductivity should be {\it universally} observed
in 2DES in a magnetic field, as long as inequalities 
(\ref{eq:omegac>T}) and (\ref{eq:omegacgamma}) are fulfilled.
For example, 
in a Si inversion layer sample \cite{wakabayashikawaji}
with $m\sim 10^{-1}m_0$ ($m_0$: free electron mass), $n\sim 10^{11}\text{cm}^{-2}$, 
and the mobility 
$\mu
=e \tau/m
=\omega_{\rm c}\tau/B\sim10^4\text{cm}^2/\text{Vs}$,
%
%
should show relevant FDT violation 
in a high magnetic field $B\gtrsim 2\text{T}$ and at low temperature 
$T\lesssim 20\text{K}$.

The caveat is that $S_{xy}$, rather than $\tilde{S}_{xy}$, should be compared with $\sigma_{xy}$, to let the FDT (in the frequency domain)
valid in the classical systems \cite{SF2017}.
If the spectrum analyzer yields $\tilde{S}_{xy}$, one needs to convert it to $S_{xy}$ using Eqs.~(C.3)-(C.6) in Ref.~\cite{SF2017}.

The FDT has been widely utilized to estimate the fluctuations (noises) from the LRs when designing, e.g., electric circuits \cite{electricalengineering1,electricalengineering2,ex3}, optical devices  \cite{ex1,ex2,ex3}, and gravitational-wave detectors \cite{ex4}.
However, we should be careful in the quantum system at low temperature.
The FDT can severely underestimate the fluctuations, if it is applied to the dissipationless components of LRs.
This is true even at $\omega=0$, 
where, by contrast, the FDT still exactly holds for the dissipative components of LRs.


{\em Summary.---}
We have studied the FDT as a relation between the observed LR functions and the observed fluctuations in a macroscopic electron system, assuming that the measurements are as ideal as possible and that the system is close to equilibrium.
It is found that the FDT can be violated even at $\omega=0$ by a macroscopically large magnitude, 
regarding the dissipationless components of LRs of 'current'.

\begin{acknowledgments}
We thank N. Shiraishi and H. Hakoshima for discussions.
This work was supported by The Japan Society for the Promotion of Science, KAKENHI No. 15H05700 and 17K05497.
\end{acknowledgments} 



\end{document}